\documentclass[12pt,manuscript]{aastex} 

\usepackage[varg]{txfonts}
\usepackage{graphics,graphicx}
\usepackage{amssymb} 
\usepackage{natbib}
\usepackage{hyperref}

\newcommand{\GS}{GS~1826$-$24}
\newcommand{\src}{GS~1826$-$24}

\newcommand{\I}{{\it INTEGRAL}} 
\newcommand{\N}{{\it NuSTAR}} 
\newcommand{\Sw}{{\it Swift}}
\newcommand{\xte}{{\it RXTE}}
\newcommand{\chandra}{{\it Chandra}}
\newcommand{\xmm}{{\it XMM-Newton}}

\newcommand{\ergcs}{erg~cm$^{-2}$~s$^{-1}$} 
\newcommand{\ergcm}{erg~cm$^{-2}$} 
\newcommand{\ergps}{erg~s$^{-1}$}

\slugcomment{Draft of \today; for submission to ApJ}

\shorttitle{Swift and NuSTAR observations of GS~1826-24} 
\shortauthors{Chenevez, Galloway, in 't Zand, et al.} 

\begin{document} 
\title{A SOFT X-RAY SPECTRAL EPISODE FOR THE CLOCKED BURSTER, GS~1826$-$24 AS MEASURED BY Swift AND NuSTAR}

\author{J.~Chenevez\altaffilmark{1},
 D.K.~Galloway\altaffilmark{2,3},
 J.J.M. in 't Zand\altaffilmark{4,5},
 J.A.~Tomsick\altaffilmark{6},
 D.~Barret\altaffilmark{7},
 D.~Chakrabarty\altaffilmark{8},
 F.~F\"urst\altaffilmark{9},
 S.E.~Boggs\altaffilmark{6},
 F.E.~Christensen\altaffilmark{1},
 W.W.~Craig\altaffilmark{6,10},
 C.J.~Hailey\altaffilmark{11},
 F.A.~Harrison\altaffilmark{9},
 P. Romano\altaffilmark{12},
 D.~Stern\altaffilmark{13}, \and
 W.W.~Zhang\altaffilmark{14}
}

\email{jerome@space.dtu.dk}

\altaffiltext{1}{DTU Space - National Space Institute, Technical University of Denmark, Elektrovej 327-328, 2800 Lyngby, Denmark}
\altaffiltext{2}{School of Physics \& Astronomy, Monash University, Clayton, VIC 3800, Australia}
\altaffiltext{3}{also Monash Centre for Astrophysics, Monash University}
\altaffiltext{4}{SRON Netherlands Institute for Space Research, Sorbonnelaan 2, 3584 CA Utrecht, The Netherlands} 
\altaffiltext{5}{Astronomical Institute, Utrecht University, PO Box 80 000, 3508 TA Utrecht, The Netherlands}
\altaffiltext{6}{Space Sciences Laboratory, University of California, Berkeley, CA 94720, USA}
\altaffiltext{7}{Institut de Recherche en Astrophysique et Plan\'{e}tologie, 9 Avenue du Colonel Roche, 31028 Toulouse, France}
\altaffiltext{8}{Kavli Institute for Astrophysics and Space Research, Massachusetts Institute of Technology, 70 Vassar Street, Cambridge, MA 02139-4307, USA}
\altaffiltext{9}{Cahill Center for Astronomy and Astrophysics, California Institute of Technology, Pasadena, CA 91125, USA}
\altaffiltext{10}{Lawrence Livermore National Laboratory, Livermore, CA 94550, USA}
\altaffiltext{11}{Columbia Astrophysics Laboratory, Columbia University, New York, NY 10027, USA}
\altaffiltext{12}{INAF-IASF Palermo, Via Ugo La Malfa 153, 90146 Palermo, Italy}
\altaffiltext{13}{Jet Propulsion Laboratory, California Institute of Technology, Pasadena, CA 91109, USA}
\altaffiltext{14}{NASA Goddard Space Flight Center, Greenbelt, MD 20771, USA}

\begin{abstract} 
We report on \N\ and \Sw\ 
observations of a soft 
state of the neutron star low-mass X-ray binary \src,
commonly known as the ``clocked'' burster. 
The transition to the soft state was recorded in 2014 June through an 
increase of the 2--20~keV source 
intensity measured by {\it MAXI}, simultaneous with a 
decrease of the 
15--50~keV intensity measured by \Sw/BAT. 
The episode lasted 
approximately two months, after which the source returned to its usual hard state. 
We analyze the broad-band 
spectrum measured by \Sw/XRT and \N, 
and estimate the accretion rate during the soft episode to be $\approx13\%\,\dot{m}_{\rm Edd}$, within the range of previous observations.
However, the best fit spectral model, 
adopting the double-Comptonization used previously, 
exhibits significantly softer components.
We detect seven type-I 
X-ray bursts, 
all significantly weaker (and with shorter rise and decay times) than 
observed previously.
The burst profiles and recurrence times vary significantly, 
ruling out 
the regular bursts that are typical for this source. 
One burst exhibited photospheric radius expansion, 
and we estimate the source distance as $(5.7\pm0.2)~\xi_b^{-1/2}$ kpc, where $\xi_b$ parameterizes the possible anisotropy of the burst emission.
Interpreting the 
soft state as a transition from an optically thin inner flow to an optically thick flow passing through a boundary layer, 
as is commonly observed in similar systems, 
is contradicted by the lower optical depth measured for the 
double-Comptonization model we find for this soft state.
The effect of a change in disk geometry on the burst behavior remains unclear.
\end{abstract}

 
\section{Introduction} 
\label{sec:intro}
Type I X-ray bursts 
arise from
unstable thermonuclear burning on the surface of accreting 
neutron stars (NSs) in low-mass X-ray binaries \citep[LMXBs; see, e.g.,][for a review]{Lew93}.
While 
variations in burst properties from source to source, and with time, are explained by changes in the accretion rate and the fuel composition at ignition,
a detailed physical understanding of most X-ray burster systems is still lacking.
For example, there is no explanation for the decrease in burst rates (leading apparently to a transition to stable burning) that 
occurs at accretion rates about a factor of ten below the theoretically expected value \cite[see, e.g.,][]{cor03,Gal08}. 
The details of the relationship between accretion rate, burning physics, burst morphology, and 
burst recurrence times are complex and still not understood.

For the majority of burst sources, that accrete a mix of hydrogen and helium from their companion, a general picture of bursting behavior  
arises with four burning regimes marked by increasing local 
accretion rate ($\dot{m}$) per NS unit area \citep[see, e.g.,][for details]{Fuji81, StrohBil06}. 
As identified in \cite{Fuji81},
case 3 burning occurs at low ($\lesssim0.01\ \dot{m}_{\rm Edd}$\footnote{Defined as the mass accretion rate, $8.8\times10^4\ {\rm g\,cm^{-2}\,s^{-1}}$, corresponding to the Eddington luminosity in a $1.4 M_\odot$ NS frame.}) accretion rates, and arises from unstable hydrogen ignition in a mixed H/He environment. No example of this type of burning has confidently been observed. At higher accretion rates, corresponding to case 2, steady burning of hydrogen commences, while helium burning is still unstable. However, $\dot{m}$ is low enough that the accreted hydrogen is exhausted at the base of the fuel layer by the time unstable helium ignition is triggered, so case 2 bursts should occur in a He-rich environment, and with relatively long recurrence times. The resulting burst light curves exhibit short ($<1$~s) rises and tails ($\lesssim10$~s), with high peak luminosities typically exceeding the Eddington limit, which generates photospheric radius-expansion. Hence, the ratio, defined as the $\alpha$ value, of the persistent fluence between bursts to the burst fluence often exceeds 100. At higher accretion rates (above a few percent of $\dot{m}_{\rm Edd}$) the burst recurrence time becomes short enough that hydrogen remains in the base of the fuel layer at ignition, and these case 1 bursts exhibit long profiles characteristic of $\beta$-decay-mediated hot-CNO burning, and $rp$-process tails. The $\alpha$ values are consistently lower than in case 2. Finally, at accretion rates $\gtrsim\dot{m}_{\rm Edd}$, He-burning should stabilise, and no further bursts are expected.

\src\ 
\citep[aka Ginga~1826$-$238, the ``clocked'' or ``textbook'' burster; see][]{ubertini}
demonstrates the closest agreement with theoretical model predictions among the 
over 100-known thermonuclear burst sources\footnote{\url{http://burst.sci.monash.edu/wiki/index.php?n=MINBAR.SourceTable}}. 
It has exhibited regular bursting behavior with highly consistent  properties from burst to burst over the 30 years since its discovery as a new transient 
(Tanaka et al. 1989). 
Indeed, using \xte\/ observations of 24 bursts, \cite{Gal04} measured a relationship between persistent X-ray flux and burst recurrence time: the latter decreases almost linearly as the accretion rate increases. 
This implies that the accreted mass between two bursts is each time completely burned during a burst, and is approximately the same even as the accretion rate changes.
The burst light curves and properties of \src\ have also been shown to be in good agreement with the predictions of time-dependent {\sc Kepler} \citep{Weaver} model predictions \cite[see][]{Heger07}. The observation-model comparisons indicate that the source normally undergoes rapid proton ($rp$)-process burning of mixed H/He fuel with approximately solar composition \cite[i.e. Case 1 of][]{Fuji81}.

Subsequent analysis of a more extensive burst sample showed 
deviations from the previously tight correlation between the flux (measured above 2.5~keV by \xte) and recurrence time. 
However, simultaneous \chandra\/ and \xmm\/ observations indicated that these deviations may result from underestimates of the persistent flux arising from  a partial redistribution to lower energies, such that the accretion rate--recurrence time relationship remained close to that expected theoretically \cite[]{Thompson_Gal}.

Due to the absence (so far) of Eddington-limited bursts, the source distance has been constrained in a variety of ways. A lower limit of 4~kpc was estimated from optical measurements \citep{barret00}, while the peak flux of sub-Eddington bursts implies an upper limit of 8~kpc \citep[]{intZ99, Kong}.
By matching the observed burst profiles with {\sc Kepler} numerical model predictions, \cite{Heger07} estimated a distance of 
$(6.07\pm 0.18)~\xi_b^{-1/2}$~kpc, where $\xi_b$ is the burst emission anisotropy factor. \citet{Zamfir} analyzed the same \xte\/ data as \cite{Heger07} to establish mass and radius constraints, as well as an upper limit on the distance of $5.5\times~\xi_b^{-1/2}$~kpc.
For these constraints (and also for this paper) the convention of \citet{fuji88} has been adopted, which defines $\xi_b$ (and the corresponding value for the persistent emission, $\xi_p$, which may have a different value) such that the luminosity $L_{b,p} = 4\pi d^2 \xi_{b,p} F_{b,p}$. Thus, $\xi_{b,p}>1$ implies that emission is preferentially beamed {\em away} from the line of sight, so that the isotropic luminosity implied from the flux measurements is an underestimate.

Since its discovery, \GS\ has consistently been observed in a persistent ``hard'' spectral state characterized by a dominant power-law component. Other burst sources are known to switch between hard and soft states, the latter associated with higher accretion rates, which last for days to months, and are accompanied by changes in burst behavior. Due to the pattern described by these sources in an X-ray color-color diagram, these states are known as the ``island'' and ``banana'' states \citep[see also Galloway et al. 2008]{vanParadijs}.
In NS-LMXBs, spectral state transitions are thought to involve variations in the accretion flow through a truncated, optically thick and geometrically thin disk. In the low-hard (island) state the accretion disk inner radius is limited by a hot optically thin quasi-isotropic inner flow, while in the high-soft (banana) state, the hot flow vanishes as the disk inner radius extends down to the NS surface, meeting a boundary layer \citep[see][]{barret01, done07}.
These changes in accretion flow geometry are related to changes in mass accretion rate, and are thought to affect the burst behavior. 
As an example, the transient X-ray burster IGR~J17473-2721 was observed in outburst in 2008, experiencing a remarkable switch from hard to soft state accompanied by a dramatic change in burst behavior, which demonstrated a hysteresis in the burst rate as a function of persistent bolometric flux \citep[see][]{Chenevez11}.
Another particular effect of the accretion flow on the burst behavior is the interaction, during the soft state, of the boundary layer with the NS atmosphere, which influences the spectral evolution of the burst emission in a way that is not observed during hard state bursts \citep[see][and references therein]{Kajava}.

On 2014 June 8, \GS\ was detected for the first time 
in a soft spectral state \citep[][see also Asai et al. 2015]{Nakahira}, which lasted more than two months, 
according to the long-term monitoring by the {\it MAXI} Gas Slit Camera \citep[GSC;][]{Matsuoka} and the \Sw\ Burst Alert Telescope \citep[BAT;][]{Krimm}. 
Here we present analysis of \N\ and \Sw\ target-of-opportunity (ToO) observations of \GS\ triggered in response to this unprecedented episode.

\section{Observations and data Analysis}
\label{data}

\subsection{MAXI}
The 
{\it Monitor of All-sky X-ray Image} \citep[{\it MAXI};][]{Matsuoka} 
has been deployed aboard the International Space Station since August 2009. 
We use publicly available data\footnote{\url{http://maxi.riken.jp/top/index.php?cid=1&jname=J1829-237\#lc}} from {\it MAXI}/GSC 
to examine the long-term 
2--20~keV intensity of \GS. 
We converted the observed GSC count-rate to mCrab units adopting 
$3.3\pm0.1\ {\rm count\,cm^{-2}\,s^{-1}}$ for 1~Crab\footnote{Equivalent to a flux of $(3.2\pm0.1)\times10^{-8}$~\ergcs\ (2--20 keV).}, 
as obtained from the average GSC count-rate 
over the same time interval between October 2013--October 2014.

\subsection{Swift}
\label{Swift}
We utilize daily averaged 
15--50~keV intensity measurements for \GS, measured by BAT \citep{Barthelmy05} 
on the \Sw\ satellite \citep{Gehrels04} downloaded from the website\footnote{\url{http://swift.gsfc.nasa.gov/results/transients/Ginga1826-238/}} 
\citep{Krimm} for this analysis.
A long-term light curve was extracted over the same time interval as for the {\it MAXI}\/ data.
The BAT count-rate was converted to mCrab adopting 1 Crab\footnote{Equivalent to a flux of $(1.5 \pm0.1)\times10^{-8}$~\ergcs\ (15--50 keV).} $= 0.22\pm0.008\ {\rm count\,cm^{-2}\,s^{-1}}$.

The \Sw\ X-Ray Telescope \citep[XRT;][]{Burrows},
which is sensitive to X-ray photons in the  0.2--10~keV band,
observed \GS\ on 2014 June 20 for 1~ks  
as a follow-up to the report of the  soft state \citep{Nakahira}.
On June 24 we requested a longer ToO observation 
with the 
goal of detecting X-ray bursts, obtaining an additional exposure of 17~ks.
A third observation was scheduled to coincide with our \N\ ToO 
(see below) on June 27, for 1.5~ks. 
All these XRT observations (see Table \ref{table:obs_log} for details) were executed in window timing (WT) mode.
The raw data were first reduced using the online XRT products  tool \citep{evans09} 
provided by the \Sw\ team at the University of Leicester\footnote{\url{http://www.swift.ac.uk/user_objects/}},
and with our own analyses, which gave consistent results with the former.
Our analyses, which are used in the present paper, were performed with standard software within HEASOFT v6.16 and CALDB files from 2014 June 10.

\begin{deluxetable}{llllccc}
\tablewidth{0pc}
\tablecaption{Log of \Sw\ and \N\ observations of \src\ in 2014 June \label{table:obs_log}}
\tablehead{
 &  & &  & \colhead{Time range} & \colhead{Exposure} & \colhead{No.} \\
\colhead{Date}    &  \colhead{MJD} & \colhead{Instr.} & \colhead{Obs. ID} & 
  \colhead{(UT)} & \colhead{(ks)} & \colhead{bursts}
}
\startdata
2014 June 20 & 56828 & \Sw/XRT & 00035342005 & 18:53–-19:10 & 0.982 & \nodata \\
2014 June 24 & 56832 & \Sw/XRT & 00035342006 & 14:06–-00:19\tablenotemark{a} & 16.65 & 1\tablenotemark{b} \\
2014 June 27 & 56835 & \N & 80001005002 & 15:36–-23:30 & 13.2 & 1 \\

 &  & \Sw/XRT & 00080751002 & 22:34-–00:13\tablenotemark{a} & 1.501 & \nodata \\
 &  & \N & 80001005003 & 23:30-–22:30\tablenotemark{a} & 38.7 & 5 \\
\enddata
\tablenotetext{a}{End time is on the following day.}
\tablenotetext{b}{The peak count-rate of the \Sw/XRT burst is $\simeq155$ count~s$^{-1}$.}
\end{deluxetable}

\subsection{NuSTAR} 
\label{sec:nustar}
The {\it Nuclear Spectroscopic Telescope Array} \citep[\N;][]{Harrison} 
consists of two identical telescopes with a 10~m focal length, focusing X-rays between 3--79 keV using depth-graded multi-layer grazing incidence optics. At the focus of the telescopes are Focal Plane Modules A and B (FPMA and FPMB), each consisting of a grid of four CZT sensors, with 32 $\times$ 32 pixels.

\N\ performed a ToO observation of \GS\ on 2014 June 27 and 28 for a total elapsed time of 108~ks divided into two contiguous datasets with exposures of 13.2~ks and 38.7~ks, respectively (Table \ref{table:obs_log}).
The \N\ data were reduced using the standard NuSTARDAS pipeline v1.4.1 utilizing CALDB files from 2014 October 20.
Images obtained from FPMA and FPMB in each dataset were used to define source and background extraction regions, both situated on the same  pixel sensor of the detector.
Light curves and spectra of \GS\ were extracted using the {\it FTOOLS} {\tt``nuproducts''} from a region of $100\arcsec$ 
radius centered on the source location in each module. 
Based on the \N\ point spread function (PSF), this aperture contains 99\% of the source counts. Another circular region of $120\arcsec$ radius centered about $280\arcsec$ from the source was used to measure the sky and instrument backgrounds.
The background outside the source extraction region is negligible ($<1\%$ of the source counts) 
below 30~keV.
For analysis of the persistent emission, we subtracted the full-bandwidth background spectrum, extracted over the same time interval as the source spectrum.

\subsection{{\it INTEGRAL}/JEM-X and {\it RXTE}/PCA data from MINBAR}

In this paper we utilize preliminary data from the Multi-INstrument Burst ARchive (MINBAR\footnote{\url{http://burst.sci.monash.edu/minbar}}).
They consist of analyses of all bursts detected in public \xte/PCA \citep{Jahoda} and {\it BeppoSAX}/WFC \citep{Boella} data through the whole lifetimes of these missions, 
as well as all public data from the JEM-X camera \citep{lund03} onboard the \I\/ satellite \citep{w03}, through 2014 December.
Analysis products include full-range light curves at 0.25~s (1~s) time resolution, for \xte\/ (\I), as well as time-resolved spectral analyses following the procedures described by \cite{Gal08}.

\subsection{Time-resolved spectral analysis} 
\label{ssubsec:burstspec} 

We extracted time-resolved spectra covering each burst observed with \Sw\ and \N\, and carried out spectroscopy on these data as follows. We first defined time bins using full-energy range light curves at 0.25-s time resolution. We subtracted the pre-burst level, and defined time bins forward and backward from the time of peak count-rate such that each bin had approximately the same number of detected counts. For \Sw, the aim was 350~counts; for \N, the aim was 200~counts each in FPMA/B. The shortest time bin for the \Sw\ burst was 3~s; for the \N\ bursts, 1~s. Half of the time bins for the \N\ bursts were 3~s or shorter.
Trial-and-error suggests that shorter bins offer no improvement on the spectral fit parameters.

The \N\ burst data were significantly affected by dead time, as is commonly the case when observing bright objects \citep{Harrison}. 
This effect reduces the detected count-rate below that incident on the detectors, and so a correction must be applied \citep[see][]{Bachetti}. 
The most energetic burst (\#3; see section \ref{subsec:bursts}) reached a peak net intensity of 
approximately 1200~count~s$^{-1}$, which corresponds to almost twice the Crab count-rate (corrected for dead time, PSF, and vignetting).
At this intensity, and including the pre-burst (persistent) emission, the dead time fraction was about 0.75. At the median count-rate for all the bins of 200~count~s$^{-1}$, the dead time fraction was 0.4. The high dead time fraction necessitated the time binning described above being performed on the detected counts (rather than the inferred incident count-rate).

We rebinned each spectrum to ensure at least 10 counts per bin.
We fit each spectrum with an absorbed blackbody model, with the neutral absorption fixed at $4\times10^{21}\ {\rm cm^{-2}}$ 
(in 't Zand et al. 1999). 
For the \Sw\ spectra, we fit in the range 0.3--10~keV and included a systematic error of 3\%, as recommended in the \Sw\ CALDB release note \#9\footnote{\url{http://www.swift.ac.uk/analysis/xrt/files/SWIFT-XRT-CALDB-09\_v16.pdf}}.
For the \N\ spectra, we assumed no systematic error and fit in the energy range 3--20~keV. 

\section{Results}  
\label{sec:results}

\label{sec:longterm} 
\begin{figure}[hp] 
\includegraphics[angle=-90, width=15cm]{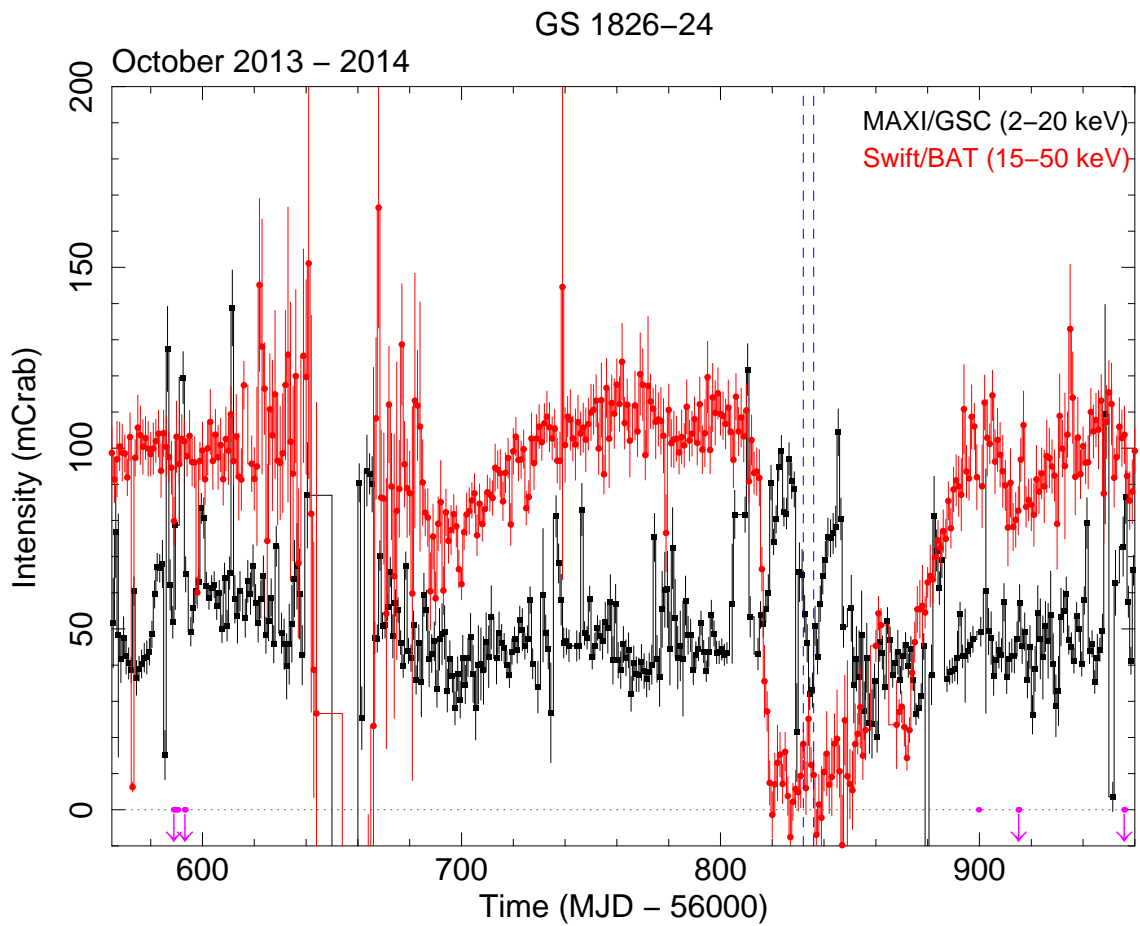}
\caption
{Daily averaged persistent intensity  of \GS\ between 2013 October 1 and 2014 October 30 as measured by {\it MAXI} and \Sw/BAT.
The data gap  between MJD 56640 and 56665 corresponds to the time the source could not be observed due to instrumental sun-angle constraints.
The time interval of our \Sw/XRT and \N\ observations is indicated by vertical dashed lines (MJD 56832 -- 56836).
Arrows on the time axis indicate the dates of the  bursts detected by {\it INTEGRAL}/JEM-X (see section \ref{burstregime}), and the corresponding observation coverage is shown on the horizontal 0-line.
} 
\label{fig:pers} 
\end{figure}

We show the long term intensity and spectral state history of \GS\ in Fig. \ref{fig:pers}, 
via the joint {\it MAXI} and \Sw/BAT light curves, where the times of the \N\ and \Sw\ ToO observations are indicated.

Beginning around MJD~56803 (2014 May 26) the 2--20~keV {\it MAXI}/GSC intensity increased over a week-long interval to more than a factor of two higher than the typical value of 45~mCrab. 
During this excursion, the 15--50~keV \Sw/BAT intensity was steady. 
A closer inspection of the {\it MAXI} light curve at the orbital resolution reveals that the flare on MJD~56810 was likely due to an X-ray burst, and a handful more bursts were detected by {\it MAXI} all along the source soft state episode. 
One of these bursts occurred during the time interval covered by our observations, but unfortunately at a time coinciding with one of the \N\ orbital data gaps (see \S \ref{subsec:bursts}); the other bursts were separated by more than one day from our \Sw\ and \N\ observations.
The {\it MAXI}/GSC daily intensity returned to the pre-flare level by MJD~56814, but one day later began to increase again, this time accompanied by a steep decrease in the \Sw/BAT intensity. On MJD~56820 (2014 June 12), the 15--50~keV intensity became below the \Sw/BAT detection level, and remained below 20~mCrab for the next 30 days.
The 2--20~keV intensity was above the typical level through to MJD~56850, excluding a 5-day interval beginning MJD~56832 (and coinciding with the scheduling of our ToO observations). The 15--50~keV intensity recovered to the typical level of approximately 110~mCrab over a much more extended period of about 50 days.
For the sake of completeness
\footnote{At the time of writing this paper, we note that a similar soft episode of \src\ was recorded by BAT and {\it MAXI} for a duration of about 20 days around 2015 June 3 (MJD~57176), and again from 2015 July 9 (MJD~57212) through August 2015.}, 
we note a previous episode in 2013 October-November during which the {\it MAXI} daily light curve seems to make a few short excursions to approximately the same level as in 2014 June, although the BAT count-rate did not simultaneously decrease \citep[see\,also][Fig. 2]{Asai}.

\subsection{Persistent emission}  
\label{sec:persistent} 
We investigated the persistent spectrum using 0.3--10~keV \Sw/XRT 
and 3--78~keV \N\ 
spectra. 
It appears from the long term {\it MAXI} light curve shown in Fig. \ref{fig:pers} that our \Sw\ and \N\ observations were performed 
while the 15--50~keV intensity was still suppressed, but the 2--20~keV intensity had
temporarily returned to $\approx50$~mCrab, roughly consistent with the level prior to the flaring activity.

\begin{figure}[hp]  
\includegraphics[angle=-90, width=\textwidth]{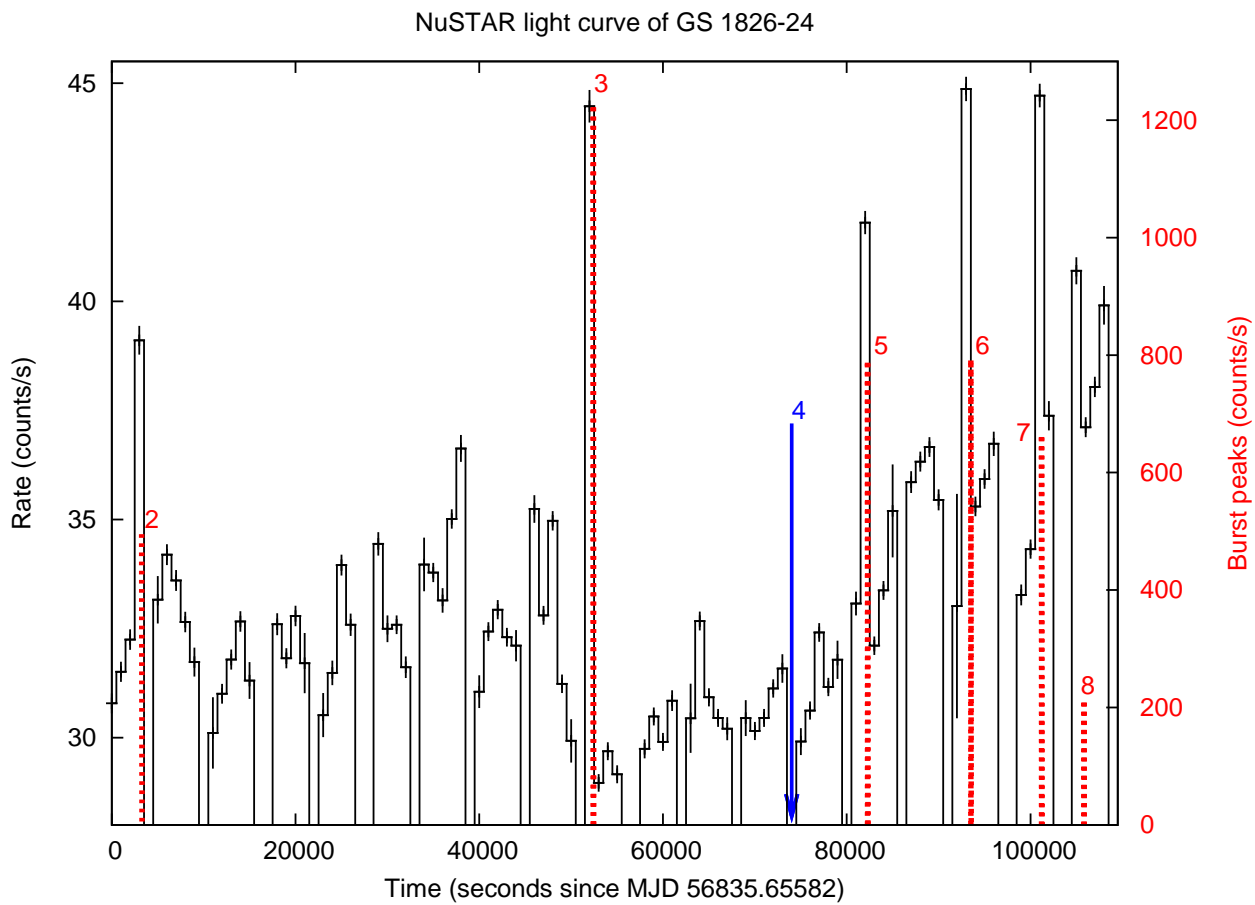}
\caption
{X-ray intensity of \GS\ measured in the 3--42~keV band by \N/FPMA during  2014 June. 
The persistent emission (at 1000~s) resolution is shown (black symbols and histogram, left-hand $y$-axis) 
along with the time of the bursts (dashed red lines); the peak intensity is indicated by the length of the lines (right-hand $y$-axis). 
The blue arrow indicates the time of the {\it MAXI} burst (\#4).
} 
\label{fig:fpma_lc} 
\end{figure} 

The source intensity light curve obtained with \N\ is shown in Fig. \ref{fig:fpma_lc}, 
where both the variation of the persistent intensity between bursts, and the  peak count-rate of six bursts are displayed simultaneously. 
Burst \#3 (see below), which has the highest peak intensity, occurred 
after the longest separation from the previous event (assuming no burst is missed during the regular data gaps). The persistent count-rate was steady at approximately 32~count~s$^{-1}$ within this interval, dropping slightly to a minimum immediately following the burst, and from  this point rising steadily to a level about 30\% higher towards the end of the observation.

In order to establish a cross-calibration of \Sw/XRT and \N/FPMA and FPMB \citep[see also][]{Madsen}, we first identified all the times of overlap between the observations with the two instruments. There were only two such intervals, between 
MJD~56835.94105 and 56835.94730 (duration 540~s), and between 
MJD~56835.99881 and 56836.00189 (duration 266~s). 
We refer to these two intervals as O1 and O2, respectively. 
We extracted \Sw/XRT spectra from observation 00080751002 over each of these intervals, and \N/FPMA and FPMB spectra from observations 80001005002 (O1) and 80001005003 (O2).

We carried out a joint fit of the spectrum for 
both intervals O1 and O2 simultaneously in the range 0.3--10~keV (XRT) and 
3--40~keV (\N),  
with the double Comptonization model adopted by \cite{Thompson_Gal}. 
We grouped the XRT and \N\ spectra to ensure a minimum of 10 counts per bin for XRT, and 30 counts per bin for \N.
No source emission was detected with \N\/ above 50~keV.
We set the neutral absorption along the line of sight with the column density frozen at $4\times10^{21}\ {\rm cm^{-2}}$ \citep{intZ99} 
with updated inter-stellar medium abundances ({\sc Xspec} {\tt tbabs} model of \citealt{Wilms}).
The composite model consists of two {\tt compTT} components in {\sc Xspec} \citep[][and references therein]{Arnaud96}, one with a low electron temperature $kT_e$ and high optical depth $\tau$, and the other with a high $kT_e$ and low $\tau$. The electron temperature for the high-$kT_e$ component was effectively unconstrained in the fits, and so we froze this value at 20~keV \citep[as measured by \xte\/ observations in 2002--3; see][]{Thompson_Gal}. 
The resulting fit, with the spectral parameters tied between the two intervals, gave a reduced $\chi^2=1.011$ for 1804 degrees of freedom.
The full set of spectral fit parameters are listed in Table \ref{table:specfits}, and the unfolded spectrum and data-to-model ratio for interval O1 are shown in 
{Figure \ref{fig:spec_compare}.

Simpler spectral models, such as a single Comptonization component, do not yield acceptable fits for plausible absorption columns.
We also included a constant multiplicative factor in order to establish any relative flux offset between the \Sw/XRT and \N\ instruments. The best-fit value of this parameter was 
$1.022\pm0.015$,
indicating that the two instruments are consistent within their absolute flux calibration.

As expected based on the \Sw/BAT and {\it MAXI} light curves, the spectrum measured by \Sw\ and \N\/ in 2014 June was substantially softer than previous measurements. The electron energy for the softer component ($kT_{e,1}$ in Table \ref{table:specfits}) was about a factor of two lower, while the optical depth $\tau_1$ was similar. Although we cannot constrain the electron temperature $kT_{e,2}$ for the second component, with that parameter fixed at roughly the same value observed previously, the optical depth for this, $\tau_2$, was less than half the previous value, indicating a spectrum decreasing much more steeply to higher energies. This is illustrated by the comparison with the most recent \xte\/ observation, on MJD~55683.59171, of 10.126~ks duration, with the Proportional Counter Units (PCUs) 1,2,4 active
(Fig. \ref{fig:spec_compare}).
At that time the hardness ratio of \Sw/BAT to {\it MAXI}\/ intensities was $\gtrsim 1.7$, 
compared to the corresponding value in 2014 June of $\lesssim 0.7$.

\begin{figure}[hp] 
\includegraphics[angle=270,width=\textwidth]{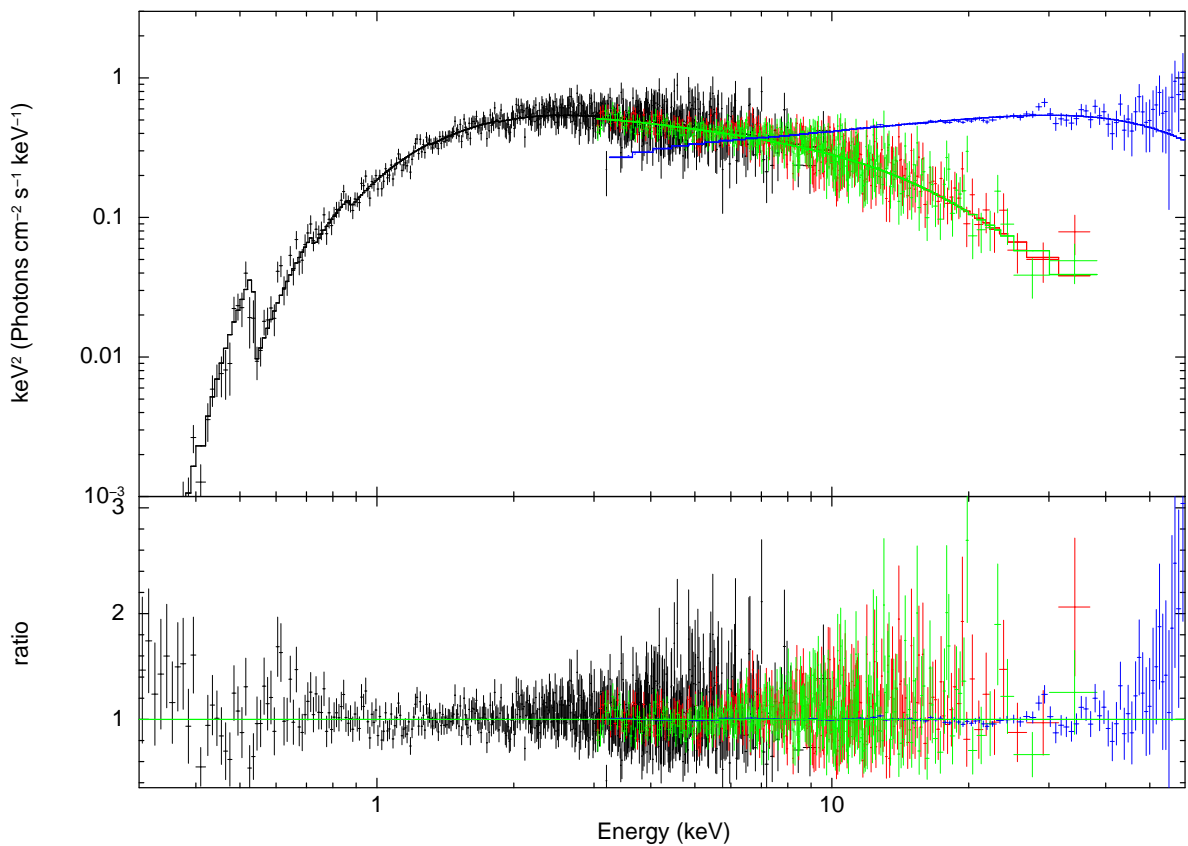}
\caption
{Persistent spectra of \Sw/XRT and \N\ data for \src\ during interval O1 of the 2014 June observation, along with the most recent \xte\ spectrum, from 2011 May 2 ({\it top panel}). The spectra are plotted as $\nu^2 S(\nu)$ to highlight the difference in spectral hardness. The black symbols (histogram) show the data (model) for the XRT data, while the red and green symbols (histogram) show the data (model) for the \N\ FPMA and FPMB data. No source emission was detected by \N\ above 50~keV. 
The blue symbols (histogram) show the spectrum observed by \xte\ on MJD~55683.59171 (obs ID 96306-01-01-03, data only from PCU \#2 shown), when the source was still in its hard state. The fitted model for the XRT and \N\ data are the best-fitting double-Comptonization model; for \xte, a single Comptonization component is used. 
The lower panel shows the data-to-model ratio for the best-fit model with parameters listed in Table \ref{table:specfits} for the XRT and \N\ data.} 
\label{fig:spec_compare} 
\end{figure} 

We then applied the double Comptonization model to each of the inter-burst intervals for bursts \#2--8 (see \ref{subsec:bursts}). For the interval between bursts 2 and 3, which spans the two \N\ observations, we extracted for simplicity a spectrum only from observation 8001005003, because the average count-rate was about the same; this covers 6.81~hr of the total (13.636~hr) separation.
We fitted these spectra simultaneously with the double Comptonization model, and experimented by trial-and-error, allowing different combinations of parameters to vary between the intervals. 
We first freed each of the Comptonization normalizations, and found that freeing only one additional parameter, $\tau_2$, was sufficient to obtain an adequate fit overall, with $\chi^2_\nu=1.0305$ ($P= 0.069$) for   4806~DOF (Table \ref{table:specfits}).
We used the {\tt cflux} convolution model component in {\sc Xspec} to measure the unabsorbed model flux within each interval in the 
3--25~keV energy range. 

As is customary, we used an ``ideal'' response to extrapolate the best-fitting spectral model outside the instrument bandpass to the range 0.1--1000~keV, and adopted this as the bolometric flux \citep[see also][]{Thompson_Gal}. Unlike the previous study by \citet{Thompson_Gal}, for which the correction to the bolometric flux based on the absorption was approximately 5\%, the spectrum during the 2014 June observations was so soft that the correction was closer to 
35\%. 
We estimate the average unabsorbed bolometric flux \cite[for comparison to the results of][]{Thompson_Gal} at 
$(2.7\pm0.2)\times10^{-9}$~\ergcs.

\begin{deluxetable}{lcccccccc}
\tabletypesize{\scriptsize}
\rotate
\tablewidth{0pc}
\tablecaption{Persistent spectral fit parameters for \src\ in 2014 June
\label{table:specfits}}
\tablehead{
 Double Comptonization model  &  & \multicolumn{7}{c}{Interval\tablenotemark{a}} \\
 \colhead{Parameter} & \colhead{Units} & \colhead{O1 \& O2} 
 & \colhead{$t_2$--$t_3$} 
& \colhead{$t_3$--$t_4$}
& \colhead{$t_4$--$t_5$} 
 & \colhead{$t_5$--$t_6$} 
 & \colhead{$t_6$--$t_7$} 
 & \colhead{$t_7$--$t_8$} 
}
\startdata
$N_H$ & $10^{21}\ {\rm cm^{-2}}$ & \multicolumn{7}{c}{-------------------------------------------------------------- (4.0) --------------------------------------------------------------} \\
$kT_{0,1}$ & keV & $0.092^{+0.012}_{-0.015}$ & \multicolumn{6}{l}{------------------------------------------------ 
$0.379^{+0.018}_{-0.024}$ ---------------------------------------------------} \\
$kT_{e,1}$ & keV & $3.04^{+0.18}_{-0.16}$ & \multicolumn{6}{l}{------------------------------------------------ 
$4.2\pm0.2$ ------------------------------------------------------} \\
$\tau_1$ & & $4.79^{+0.18}_{-0.17}$ & \multicolumn{6}{l}{------------------------------------------------ 
$2.65\pm0.14$ ------------------------------------------------------}\\
$kT_{0,2}$ & keV & $0.404^{+0.016}_{-0.015}$ & \multicolumn{6}{l}{------------------------------------------------ 
$1.592\pm0.01$ -----------------------------------------------------} \\
$kT_{e,2}$ & keV & \multicolumn{7}{c}{-------------------------------------------------------------- (20) --------------------------------------------------------------}\\
$\tau_2$ & & $0.79\pm0.05$ & $0.390^{+0.016}_{-0.015}$ 
& $0.494\pm0.016$ & $0.372\pm0.018$
& $0.083\pm0.017$ & $0.064^{+0.018}_{-0.017}$ & $0.037^{+0.018}_{-0.015}$ \\
$\chi^2_\nu$ (DOF) & & 1.011 (1804) 
& \multicolumn{6}{c}{--------------------------------------------- 1.0181 (5650) ------------------------------------------------}\\
Absorbed flux (3--25 keV) & $10^{-9}\ {\rm erg\,cm^{-2}\,s^{-1}}$ & & 
$1.1646\pm0.0017$ &
$1.0727\pm0.0016$ &
$1.100\pm0.003$ &
$1.178\pm0.002$ &
$1.183\pm0.003$ &
$1.236\pm0.005$ \\
Unabsorbed flux\tablenotemark{b} (0.1--1000~keV) & $10^{-9}\ {\rm erg\,cm^{-2}\,s^{-1}}$ & & 
$2.539\pm0.004$ & 
$2.283\pm0.003$ & 
$2.381\pm0.006$ & 
$2.653\pm0.005$ &
$2.689\pm0.006$ &
$2.782\pm0.011$ \\
\enddata
\tablenotetext{a}{Time interval between bursts \#i and \#i+1.}
\tablenotetext{b}{Extrapolated, assuming an ideal response. The flux is calculated as the mean of the fluxes for the models over each of the \Sw/XRT and \N\ spectra, and the uncertainty is calculated as the standard deviation.}
\end{deluxetable}

\subsection {Thermonuclear bursts}
\label{subsec:bursts}
We detected seven type-I (thermonuclear) bursts during our ToO observations in 2014 June, the first by \Sw\ on MJD~56832.99124, and the remaining six detected by \N\ as shown in Fig. \ref{fig:fpma_lc}. 
As mentioned above, one more burst appears in the {\it MAXI} orbital light curve at $MJD~56836.5224\pm40s$ (M. Serino; private communication). We do not have more detailed data for this burst, but we note that its time coincides with a \N\ orbital data gap.
We number these bursts \#1 to 8 based on their chronological order (see also Table \ref{table:bursts_log}).

The bursts we observed 
were significantly shorter than previous bursts, as determined by the 
duration over which the count-rate exceed 25\% of the maximum.
The typical timescales are $\approx12$~s (see Table~\ref{table:bursts_log}), compared to $35.9\pm0.4$~s for the {\it RXTE}\/ bursts in the MINBAR sample 
(Fig. \ref{fig:burst_lc}).
We fitted a one-sided Gaussian to the rising part of each burst and translated the standard deviation to the time it takes the Gaussian to rise from 25\% to 90\% of the peak value   \citep[corresponding to 1.206 times the standard deviation; see also][]{Gal08}.
We observed considerable diversity among the \N\ bursts, both in burst rise time (in the range 2--12~s) and peak intensity (a range of a factor of 8).
This inconsistency between successive bursts is also atypical for this source \cite[e.g.][]{Gal04}.
The brightest \N\ burst exhibited the shortest rise time, of 2~s.
The last burst observed (\#8) was also the weakest, and occurred after the shortest recurrence time ever observed in this source (see below). 
The rise time for this burst (and also burst \#5) was similar to the decay time, so that the burst was almost symmetric in profile.

\begin{figure}[htb] 
\includegraphics[angle=0]{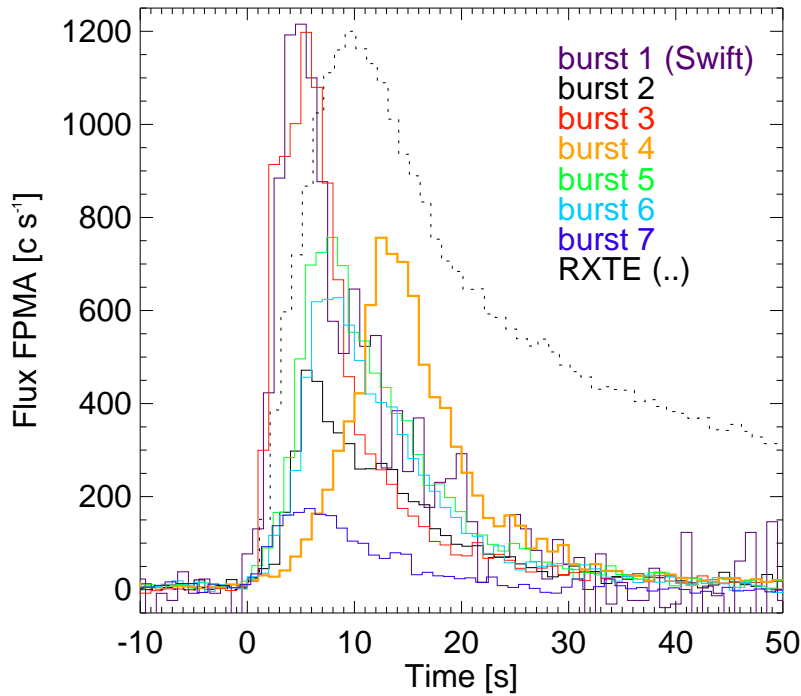}
\caption
{Light-curves of the \Sw/XRT burst (\#1; 0.2--10 keV) and the six \N\ bursts (\#2, 3, 5, 6, 7, 8; 3--42 keV) compared with a burst detected with {\it RXTE} (2--60 keV) on 1997 November 5. The pre-burst average count-rates are subtracted, and the \Sw/XRT and {\it RXTE} burst peaks are normalized to the highest \N\ peak (burst 3) at about 1200~counts~s$^{-1}$.} 
\label{fig:burst_lc} 
\end{figure} 

The shortest separation between any observed burst pair was between the last two bursts observed by \N, \#7 and 8 in Table \ref{table:bursts_log}. These events were observed on MJD~56836.82851 and 56836.88213, respectively, with a separation of 1.287~hr. Previously, \src\/ has exhibited consistently regular bursts, so we tested whether the bursts observed in 2014 June were consistent with a regular recurrence time. The separations of the previous three pairs, at 2.1, 3.179 and 2.082~hr, respectively, are not consistent with the separation for the final pair, nor any integer multiple, as expected if bursts were missed in data gaps. However, the final burst observed with \N\ was much weaker in both fluence and peak flux (see Table \ref{table:bursts_log}) than the other 
bursts in the same observation, so we consider the possibility that the final burst was the second (or third) component of a so-called ``short-recurrence time burst'', groups of up to four events seen in most sources accreting mixed H/He fuel \citep{Keek10}. Although 1.287~hr is beyond the usual range of delays seen for such events, it is possible the last burst followed more closely another event which fell in the data gap which ended just 7.81~min earlier. In that case, we should discount the final burst, and consider only the four previous ones (including the {\it MAXI} burst). The separation between the successive pairs of bursts were related in approximately a 3:2 ratio, suggesting that the bursts were occurring regularly every 1.05~hr. However, if that were the case, the expected time of one of the missing bursts between the observed events \#5 and 6 fell in the middle of an observation interval in which no bursts were observed.
Thus, we can rule out regular bursting during the 
time interval covered by our observations at high confidence.

\begin{deluxetable}{clcccccccc}
\tabletypesize{\scriptsize}
\tablewidth{0pc}
\tablecaption{Properties of thermonuclear bursts from \src\ detected in \Sw\ and \N\ observations in 2014 June
\label{table:bursts_log}}
\tablehead{
\colhead{Burst}    &   &  & 
  \colhead{Start time} & \colhead{$\Delta t$}  & \colhead{Rise time\tablenotemark{a}} & \colhead{Timescale\tablenotemark{a}} & &  \\
\colhead{no.}    &  \colhead{Instr.} & \colhead{Obs. ID} & 
  \colhead{(MJD)} & \colhead{(hr)} & \colhead{(s)} & \colhead{(s)} &
  \colhead{Peak flux\tablenotemark{b}}  & \colhead{Fluence\tablenotemark{c}} & \colhead{$\alpha$\tablenotemark{d}}
}
\startdata
1 & \Sw/XRT & 00035342006 & 56832.99124 & \nodata & $1.9\pm0.3$ & $11.3\pm1.6$ & $40_{-20}^{+80}$ & $0.21\pm0.08$ & \nodata \\
2 & \N & 80001005002 & 56835.69484 & 64.89 & $2.4\pm0.2$ & $12.3\pm1.1$ & $13.8\pm1.1$ & $0.187\pm0.006$ & \nodata\\
3 & \N & 80001005003 & 56836.26299 & 13.636 & $1.13\pm0.08$ &  $8.6\pm0.6$ & $40\pm3$ & $0.370\pm0.009$ & $337\pm8$\\
4 & {\it MAXI} & \nodata & 56836.5224\tablenotemark{e} &  $6.2\pm0.3$  & \nodata & \nodata & \nodata & \nodata & \nodata \\
5 & \N & 80001005003 & 56836.60928 & $2.1\pm0.3$  & $5.52\pm0.19$& $11.8\pm0.9$ & $25.1\pm1.9$ & $0.335\pm0.008$ & $54\pm8$\\
6 & \N & 80001005003 & 56836.74176 & 3.179  & $3.35\pm0.16$ & $12.2\pm0.9$ & $27\pm2$ & $0.352\pm0.009$ & $86\pm2$ \\
7 & \N & 80001005003 & 56836.82851 & 2.082  & $3.42\pm0.17$ & $12.6\pm1.1$ & $21.8\pm1.6$ & $0.280\pm0.007$ & $72.1\pm1.8$\\
8 & \N & 80001005003 & 56836.88213 & 1.287  & $3.4\pm0.3$ & $12.4\pm1.4$ & $6.6\pm0.6$ & $0.083\pm0.005$ & $156\pm9$\\
\enddata
\tablenotetext{a}{Measured from the count-rate burst light curve from 25\% to 90\% of the peak value in the relevant energy band.}
\tablenotetext{b}{Extrapolated peak bolometric flux in units of $10^{-9}$~\ergcs.}
\tablenotetext{c}{Integrated burst bolometric fluence in units of $10^{-6}$~\ergcm.}
\tablenotetext{d}{As every burst interval was interrupted by at least one data gap, the $\alpha$-values must formally be considered upper limits.}
\tablenotetext{e}{This burst is recorded from the public {\it MAXI} orbital light curve.}
\end{deluxetable}

\subsection{Burst energetics and spectral variations} 

We carried out time-resolved spectral analysis as described in \S\ref{ssubsec:burstspec}.
We found an adequate fit to each time-resolved net burst spectrum (with the pre-burst emission subtracted as background) using an absorbed blackbody model. 
The resulting distribution of reduced $\chi^2$ values is shown in Fig. \ref{fig:fitstats}. The maximum value for any of the fits was 1.34; this is consistent with expectations for a good fit given the number of degrees of freedom.
We list the burst spectral parameters in Table \ref{table:bursts_log}.

\begin{figure}[hp] 
\includegraphics[angle=0]{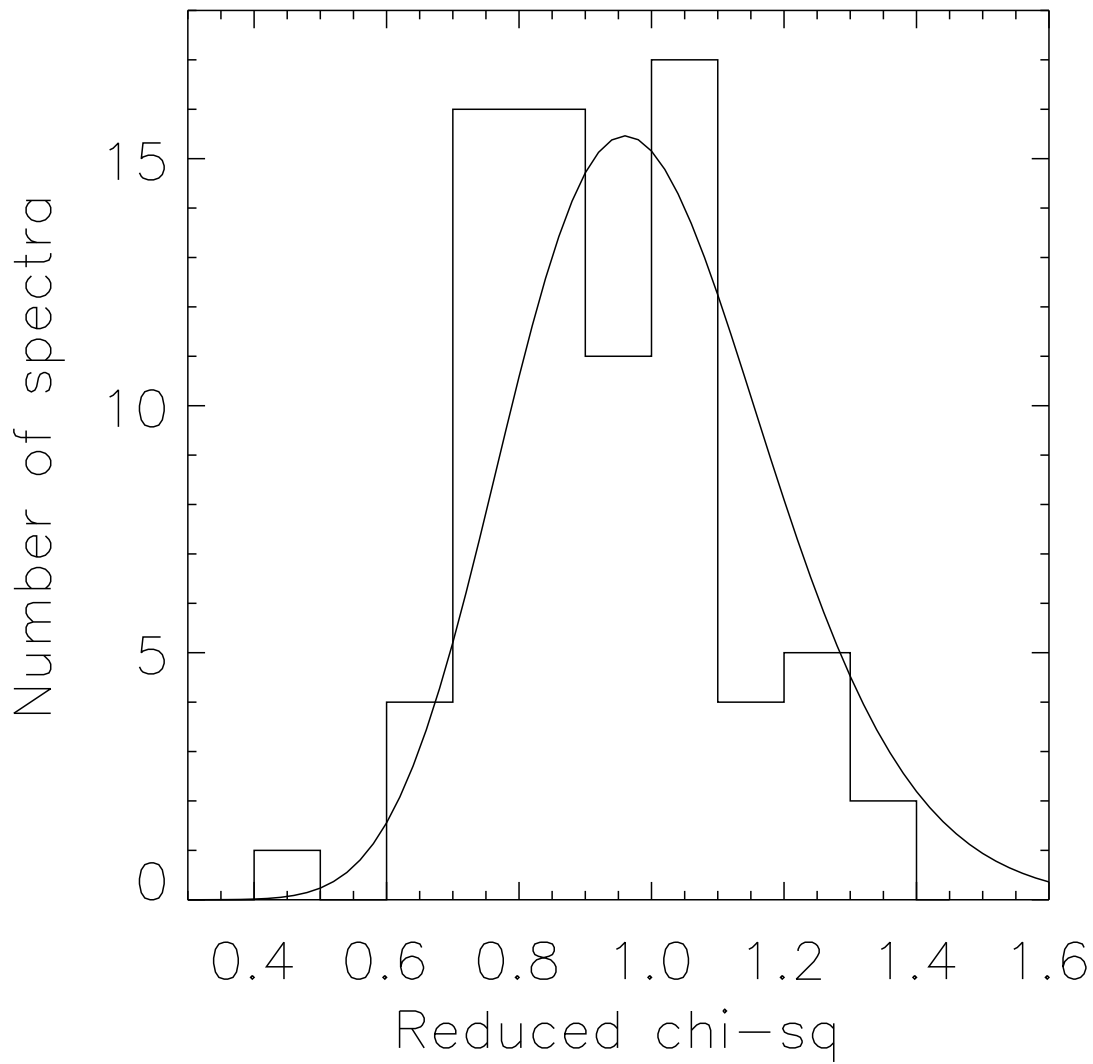}
\caption{Distribution of reduced-$\chi^2$ fit statistic for 76 spectra covering the six bursts observed by \N\ from \src. The smooth curve overplotted is the expected distribution assuming the adopted model (an absorbed blackbody with fixed neutral column density) is correct. The smooth curve is calculated for the average number of degrees of freedom in the fit (50).} 
\label{fig:fitstats} 
\end{figure} 

Figure 6 
shows time-resolved spectroscopic results for the \Sw\ and \N\ bursts. 
The time-resolved spectroscopic analysis of the brightest burst, \#3, indicates the characteristic evolution of a photospheric radius expansion (PRE) burst, with a local maximum observed in the blackbody normalization at the same time as a minimum in the blackbody temperature (Fig. \ref{fig:burst3_trs}). 
The presence of PRE in other sources is strongly correlated with the source being in a soft state \cite[]{muno04a}.

The peak flux reached during this burst  was 
$(40\pm3)\times10^{-9}$~\ergcs.
This value is a factor of 1.42 higher than the mean peak flux of the non-PRE bursts observed by \xte\/ since 2000, of $(28.4\pm1.2)\times10^{-9}$~\ergcs, and a factor of 1.24 higher than the peak flux of the brightest burst yet observed from the source: $(32.6\pm1.0)\times10^{-9}$~\ergcs as qutoed from the MINBAR database.

\begin{figure}[htb] 
\figurenum{6a}
\includegraphics[angle=-90, width=0.99\textwidth]{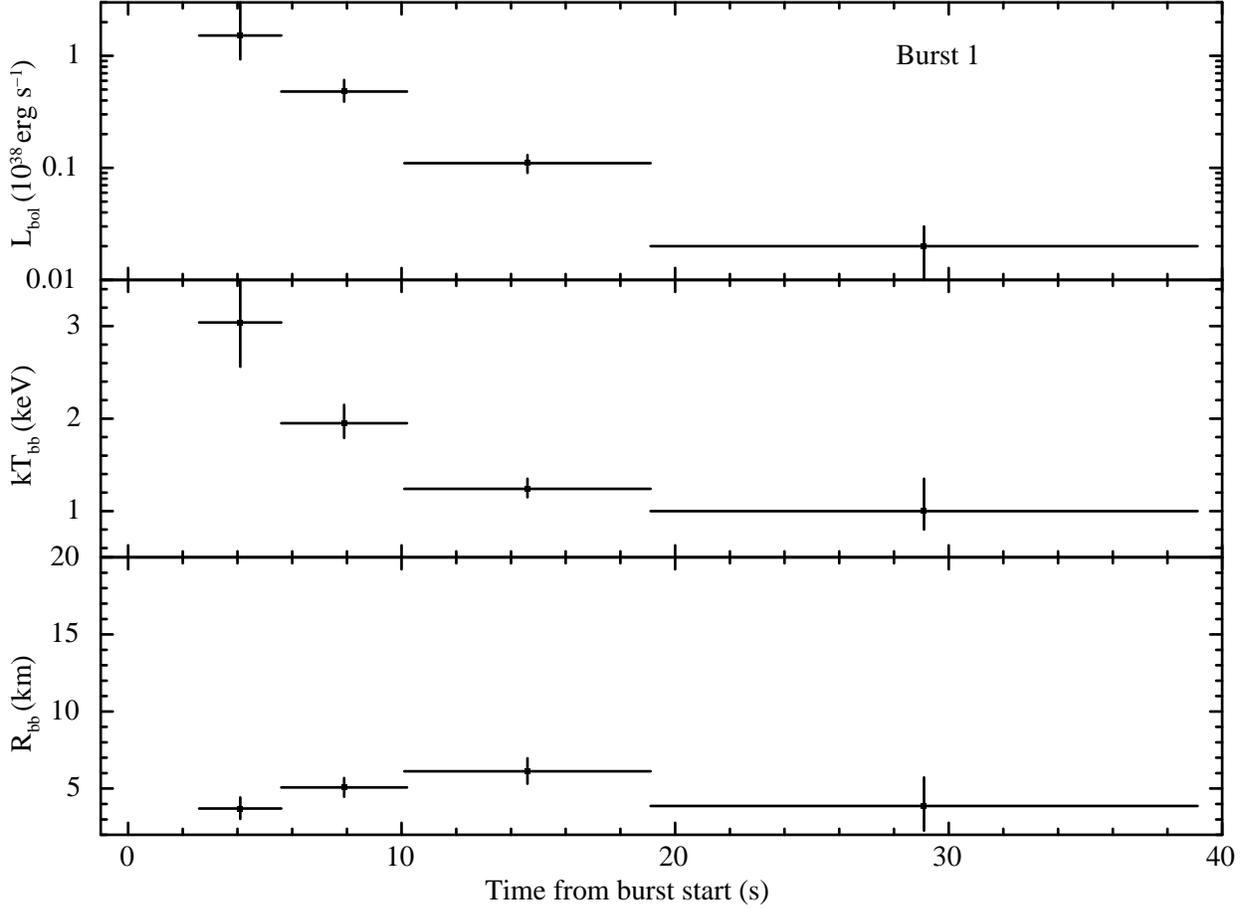}
\caption
{Time-resolved spectroscopy of burst \#1, observed by \Sw/XRT on MJD~56832.99124. The top panel shows the inferred bolometric luminosity, assuming a distance of 5.7~kpc. The middle panel shows the best-fit blackbody temperature, and the lower panel shows  the normalization.} 
\label{fig:burst1_trs} 
\end{figure} 

\begin{figure}[htb]  
\figurenum{6b}
\includegraphics[angle=-90, width=0.99\textwidth]{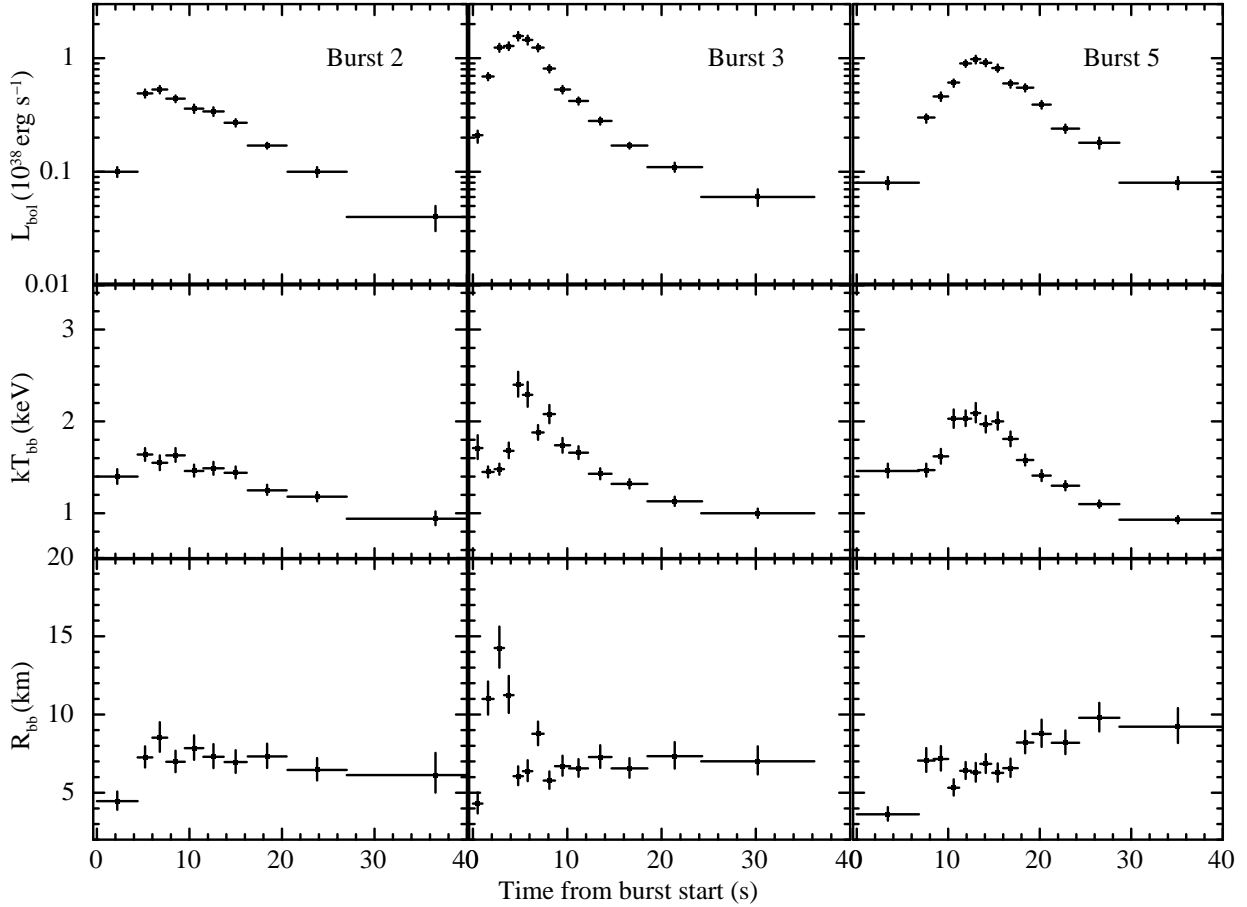}
\caption
{Same as Fig. 6a for bursts \#2, \#3, and \#5, observed by \N . Note the moderately strong radius expansion of burst \#3 during the first four seconds of the rise, coupled with a decrease in $kT_{\rm bb}$.} 
\label{fig:burst3_trs} 
\end{figure} 

\begin{figure}[htb]  
\figurenum{6c}
\includegraphics[angle=-90, width=0.99\textwidth]{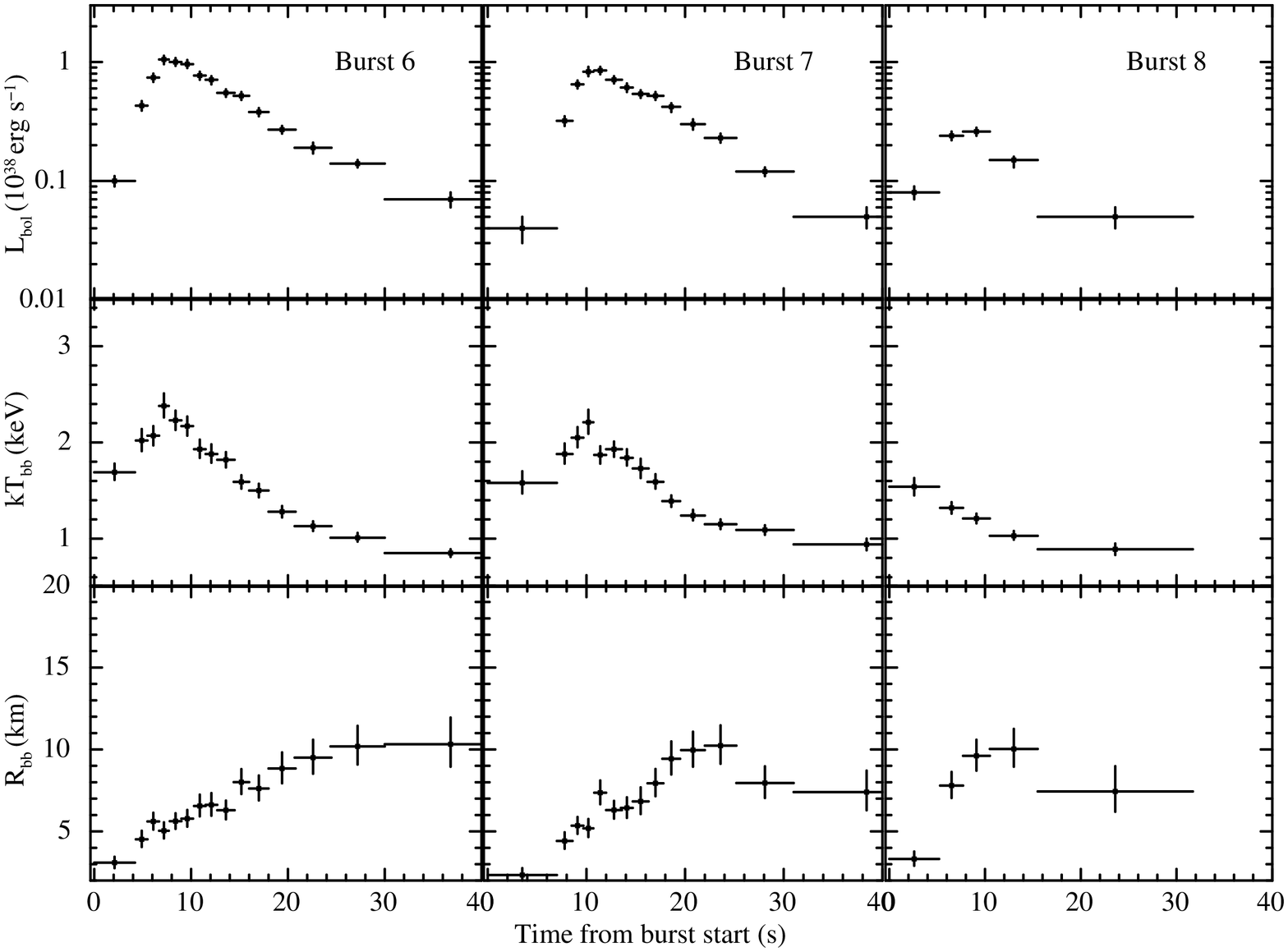}
\caption
{Same as Fig. 6b for bursts \#6, \#7, and \#8, observed by \N .} 
\label{fig:bursts_5-6-7_trs} 
\end{figure} 

We integrated over the measured fluxes to give the fluence $E_b$ for each burst, and
computed the burst timescale $\tau$ as the ratio of the 
fluence to peak flux, i.e. $\tau=E_b/F_{\rm peak}$.
A measure of burst energetics is given by $\alpha= \Delta t_{\rm rec}F_{\rm pers}/E_{b}$,
the ratio of persistent and burst fluences \citep[e.g.][]{lewin83}.
In this expression for $\alpha$, 
$F_{\rm pers}$ and $\Delta t_{\rm rec}$ are the average bolometric persistent flux and the waiting time since the last preceding burst, respectively.  
We use the {\it MAXI} burst so as to better constrain the $\alpha$-value of the following \N\ burst (\#5), although this is obtained with a relatively high uncertainty due to the approximate knowledge of the time of the {\it MAXI} burst.

\section{Discussion}
\label{sec:Discussion}

The 2014 June soft spectral state of \src\ was the first ever recorded for this well-studied source, and it revealed a number of new observational features, including 
the first burst exhibiting photospheric radius expansion, and
weak, irregular bursting behavior, including
the shortest burst interval ($1.29$~hr) measured to date.

\subsection{The source distance}

The brightest burst observed with \N, \#3, exhibited spectral evolution consistent with PRE, thought to indicate the burst flux reaching the Eddington limit.
Assuming the peak flux corresponds to the Eddington luminosity for an atmosphere with solar composition, and taking into account the effects of gravitational redshift at the surface of a $1.4\ M_\odot$, 10~km radius neutron star \cite[e.g.][]{Gal08}
the inferred distance is 
$(5.7\pm0.2)~\xi_b^{-1/2}$ kpc, where $\xi_b$ represents the possible anisotropy of the burst emission (see \S \ref{burstregime}).
In at least one other system, 4U~1636$-$536, which is thought to accrete mixed H/He fuel as assumed for \GS\ \citep[see][]{B2000, Gal04}, the effective Eddington limit is thought to be instead the higher limit appropriate for a pure He atmosphere \cite[]{Gal06}.
If we instead adopt that value, the implied distance is 
$(7.4\pm0.3)~\xi_b^{-1/2}$ kpc. Further, \citet{Kuul03} measured the Eddington luminosity for a group of LMXBs with independently known distances from their globular cluster host as $(3.79 \pm 0.15)\times10^{38}$~\ergps. Based on this value the implied distance is 
$(8.9\pm0.4)~\xi_b^{-1/2}$ kpc.

These larger distances are problematic for several reasons. First,  the  non-PRE bursts observed previously reach an average maximum flux only a factor of 1.42 lower than burst \#3, implying that the non-PRE bursts exceed the Eddington limit for mixed H/He fuel. This also seems to be the case for 4U~1636$-$536, which infrequently shows PRE bursts consistent with the H/He limit \cite[]{Gal06}. However, for \GS\ the He-derived distances also exceed the upper limit of $5.5 \times\,\xi_b^{-1/2}$~kpc derived from comparing the non-PRE burst light curves to {\sc Kepler} numerical model predictions \citep{Zamfir}.
Thus, though we cannot absolutely rule out other possibilities, we adopt a distance of $(5.7\pm0.2)~\xi_b^{-1/2}$ kpc as this is the only one that satisfies the constraints obtained by \cite{Zamfir}, and we conclude that the effective Eddington limit for \GS\ is for mixed H/He fuel.

\subsection{The persistent spectral state}
\label{persstate}

Since previous observations of \GS\ have consistently found the source in the hard (island) spectral state, we discuss here to what extent the 2014 June observation is distinct from that state. Extensive previous {\it RXTE}\/ observations of other ``atoll'' class LMXBs (so named because of their characteristic pattern in X-ray color-color diagrams) find that the hard and soft X-ray colors (defined as the ratio of counts between pairs of energy bands -- for {\it RXTE}\/, the energy bands used were 8.6--18.0 and 5.0--8.6~keV for the hard color, and 3.6--5.0 and 2.2--3.6~keV for the soft color) of these sources define an arc or a Z-shaped track \citep[e.g.][]{Gal08}. 
Unfortunately, because \GS\ was never observed by {\it RXTE}\/ to go into a soft state, it's color-color diagram is not well defined, and instead all observations cluster around a soft color value of $1.69\pm0.04$, and hard color $0.865\pm0.013$. These values, extracted from the catalog of {\it RXTE}\/ observations of \cite{Gal08}, are corrected for the PCA gain, which varied over the mission; the corresponding values prior to the gain correction for the epoch closest to the end of the mission would be 15\% lower in soft color, and 4\% higher in hard. 

We estimated the corresponding colors for the 2014 June observation in the same energy bands used for the {\it RXTE} analysis, 
and for the most recent gain epoch.
We created a simulated persistent spectrum in {\sc Xspec}, adopting the best-fit 
spectral model with a response calculated for a late-epoch {\it RXTE}\/ observation. The estimated PCA colors for the source in 2014 June are 1.189 (0.476) for soft (hard) color. In other atoll sources, a significant decrease in both soft and hard color is associated with a transition to the ``banana'' or soft spectral state. Although it is impossible to be certain in the absence of a well populated color-color diagram for \GS, the spectral measurements strongly support a state transition similar to that seen in other atoll sources.
Furthermore, although the higher accretion rate that might be implied by the spectral state transition is not supported by the estimate of the bolometric flux, such discrepancies are also well-known in other atoll sources \citep[see e.g. Fig. 6 of][]{Gal08}.

For other atoll sources, the soft ``banana'' persistent spectral state is usually interpreted as indicating a higher accretion rate, and na\"ively, the higher average burst rate for \src\ during 2014 June would seem to support this interpretation. However, 
the inferred bolometric persistent flux level of 
$(2.7\pm0.2)\times10^{-9}$~\ergcs\ is in fact in the middle of the range of bolometric fluxes that the source has been observed at historically \cite[e.g.][]{Thompson_Gal}. Thus, we find no evidence to support a markedly different accretion rate, unless the radiative efficiency (or perhaps the persistent emission anisotropy) has changed markedly. The 
\Sw\ and \N\ observations 
fell between two much higher peaks of the {\it MAXI} light curve (Fig. \ref{fig:pers}); it seems likely that the source could have been up to a factor of two brighter 
still in the soft state, 
just a few days before or after.

We note that \cite{Ji14} report the diminution of the hard X-ray persistent emission during \src\ bursts observed by \xte. These authors explain such hard X-ray shortages as due to the cooling of the hot corona by the soft X-ray burst photons \citep[see also][]{Ji15}. 
We tested for similar variation in the hard X-ray emission during the six bursts observed by \N, but 
the source intensity above 30~keV was persistently so weak (only a few counts/s) that we could not find any significant variation.
This may be consistent with \citet{Ji14} results, as our observations occurred when \src\ was in a soft state during which a negligible corona or hot accretion flow is supposed to be present.

\subsection{The bursting regime}
\label{burstregime}

\GS\ has so far been characterized by consistently regular $\gtrsim100$~s long bursts recurring at approximately periodic intervals 
between 3.56 and 5.74 hr, varying inversely as an almost linear function of the source persistent flux \citep{Gal04}. 
Apart from the burst detected by {\it MAXI} on MJD~56810 (see \S\ref{sec:longterm}), we found no observations 
in the few weeks preceding or succeeding the soft episode.
However, JEM-X detected two bursts before the soft spectral episode, in 2013 October 23 and 28 (MJD 56588 and 56593),
and afterwards, on 2014 September 15 and October 25 (MJD 56915 and  56955), respectively. 
As shown in Fig. \ref{fig:pers}, the four JEM-X bursts occurred while the source spectral state was rather stable, with the (BAT / {\it MAXI}) hardness ratio consistently about $1.6~(\pm10\%$). 
All  four bursts before and after 2014 June show similar shapes and durations as other bursts from \GS\ previously observed with JEM-X in the 3--25 keV energy range, i.e. longer rise times (from 6 to 9~s) and timescales (between 40 and 60~s). 
While it is not possible to infer the burst rate with such widely separated burst detections, the long burst timescales and characteristically hard persistent spectral state strongly suggests that \GS\ was exhibiting its normal burst behavior up to 2014 June, and following.

Based on previous measurements \citep{Thompson_Gal}, at the flux level seen in 2014 June we would expect regular, consistent bursts at a recurrence time of $\Delta t\approx 4$~hr, and with $\alpha\approx30$--40. Instead, we found much weaker, inhomogeneous bursts, with fluences at most one third of the typical value measured  in the past, and correspondingly higher $\alpha$-values.
Given the lack of regularity in the bursting, and the presence of gaps between each of the burst pairs, the measured $\alpha$-values must be considered upper limits only, so we cannot rule out lower values, consistent with the usual mixed H/He burning. However, we can determine a lower limit on the amount of H in the burst, based on the assumption that all the accreted fuel is burned during the burst:
\begin{equation}
\alpha=58\left(\frac{M}{1.4 M_\odot}\right)\left(\frac{R}{10 \mathrm{ km}}\right)^{-1} \left(\frac{Q_\mathrm{nuc}}{4.4\,\mathrm{MeV\, nucleon}^{-1}}\right)^{-1}\left(\frac{1+z}{1.31}\right)\,
\left(\frac{\xi_p}{\xi_b}\right)^{-1}
\label{eq:alpha}
\end{equation}
(note that the expression in \citealt{Gal08} omits the redshift factor) where $M$, $R$ are the mass and radius of the neutron star, $1+z=(1-2GM/Rc^2)^{-1/2}$ is the surface gravitational redshift, and $Q_{\rm nuc}=1.6+4X$~MeV~nucleon$^{-1}$, where $X$ is the hydrogen fraction averaged over the burning layer.

The ratio between anisotropy for the persistent $\xi_p$ and burst $\xi_b$ emissions that appears in Eq. \ref{eq:alpha} 
has  been estimated as 1.55 for \src\ \cite[]{Heger07}. 
The modeling of \citet{fuji88} suggests that a system inclination of $75^\circ$ is required to give this value of the relative anisotropies, and further implies
that $\xi_b=1.32$ and $\xi_p=2.04$. 
These values indicate that both the burst and persistent flux are preferentially beamed away from our line of sight, and the inferred isotropic luminosities will underestimate the true values.
The combined effect on the measured $\alpha$-values will be to underestimate the true value by a factor of 1.55.

Due to the data gaps falling between each pair of bursts detected by \N, each of the $\alpha$ measurements in Table \ref{table:bursts_log} is an upper limit on the true value. 
The most constraining value should be the minimum, which is obtained 
for burst \#5, although with relatively large uncertainty due to the lack of absolutely exact timing of the {\it MAXI} burst. 
The estimated $\alpha=54\pm8$, with the uncertainty dominated by the separation from the {\it MAXI} burst, of $2.1\pm0.3$~hr.
The corresponding value of $Q_{\rm nuc}$, assuming the range of gravitational redshift $1+z=1.19$--1.28 estimated by \cite{Zamfir}, would be in the range 
2.8--3.0~MeV~nucleon$^{-1}$, implying in turn a hydrogen mass fraction at ignition of $X=0.3$--0.35.
We further note that, while low values of $X$ may arise from steady hydrogen burning prior to the burst, there has been insufficient time to reduce it to this degree. Indeed, assuming solar CNO metallicity $Z_{CNO}$, the time to burn all the hydrogen at the base of the layer is 
\begin{equation}
t_{\rm ex}=11 (Z_{CNO}/0.02)^{-1}(X_0/0.7)\ {\rm hr}
\label{eq:tex}
\end{equation}
where $X_0$ is the accreted H-fraction \cite[e.g][]{Gal04}. For solar accreted composition, there is insufficient time between bursts to reduce the average H-fraction in the fuel layer to explain the $\alpha$-value of burst \#5. 

Another way to understand the discrepancy is by
considering the column depth of  material ignited during each burst, given by:
\begin{eqnarray}
y_b&=&\frac{4\pi d^2 E_b}{4\pi R^2 Q_\mathrm{nuc}}\left(1+z\right) \\
  &=&3.0\times10^8\left(\frac{E_b}{10^{-6} {\rm erg\,cm^{-2}}}\right)
    \left(\frac{d}{10\ {\rm kpc}}\right)^2 \nonumber \\
&&\times    \left(\frac{Q_{\rm nuc}}{4.4\ {\rm MeV\,nucleon^{-1}}}\right)^{-1}
    \left(\frac{1+z}{1.31}\right)\left(\frac{R}{10\ {\rm km}}\right)^{-2}
    \xi_b\ {\rm g\,cm^{-2}}
\end{eqnarray}
again assuming implicitly that all the accreted fuel is burned in the burst.
We set the value of $Q_{\rm nuc}$ based on the assumed average fuel composition $X$ resulting from hot-CNO burning between the bursts, i.e. $X=X_0(1-0.5\Delta t/t_{\rm ex})$ (with the factor 0.5 arising because the burning takes place at the base, and we average $X$ over the entire column).
We compare this with the column depth accreted between two bursts separated by a time interval $\Delta t$, which is $\Delta M = \dot m \Delta t$, where $\dot m = \dot M/4\pi R^2$ is the mass accretion rate per surface area on the NS.
Given the estimated bolometric persistent flux $F_{pers}$ of (2.4--$2.9)\times10^{-9}$~\ergcs, and the inferred anisotropy factor, we estimate the accretion rate at 
12--15\% of the Eddington rate (at a distance of 5.7~kpc). 
This accretion rate is in the range expected for Case 1 burning of \cite{Fuji81}, corresponding to mixed H/He burning triggered by thermally unstable helium ignition. 
Assuming that the 
nuclear burning is completely conservative, one would expect $y_b$ to be close to 
$\Delta M$.
Instead, we find that $y_b$ consistently underestimates $\Delta M$, even for the relatively close pairs of bursts (Fig. 7). 

We infer that the assumption of conservative burning (i.e. that all accreted fuel is burned during the bursts) cannot be reconciled with the data, implying that some other process is reducing the available fuel prior to ignition. This burning appears to  preferentially removing  hydrogen, based on the short burst timescales and since the reduction is in excess of the normal steady hot-CNO burning.

\begin{figure}[hp] 
\figurenum{7}
\includegraphics[angle=0]{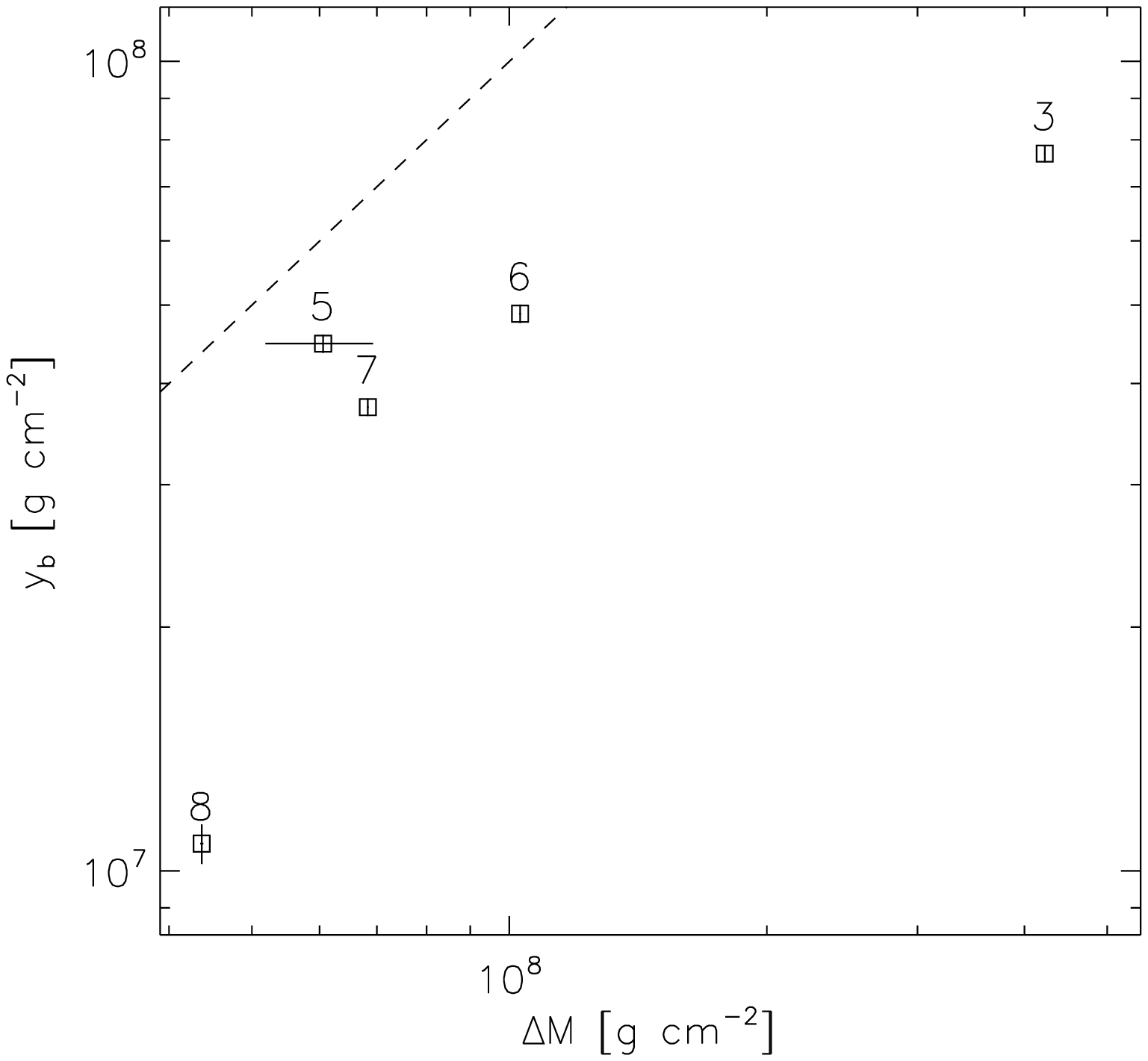}
\caption{Comparison of the estimated accreted column $\Delta M$, calculated from the inferred accretion rate and the inter-burst interval, with the ignition column $y_b$, based on the measured burst fluence and the assumed $Q_{\rm nuc}$ arising from the effects of hot-CNO burning between the bursts. Both parameters are measured in the local (neutron star) frame, assuming $1+z=1.28$. The dashed line is the line of equality, at which point the burning would be conservative. Note that the bursts consistently ignite at columns well below that accreted.} 
\label{fig:efficiency} 
\end{figure} 

We conclude that 
\GS\ 
bursts inefficiently in the soft state, 
and igniting fuel with significantly lower hydrogen fractions than the 
previously inferred solar value. The burst intervals are too short for the lower H-fractions to arise purely by hot-CNO burning between the bursts, unless the CNO metallicity is of order ten times solar. 
Steady burning of accreted fuel in addition to hot-CNO burning would explain both the relative inefficiency of the thermonuclear bursts, and would also provide an extra fuel source to explain the relatively low ignition columns.

The results of our spectral analysis of the persistent emission compared to previous observations indicate a softening, but at a similar inferred accretion rate.
Such a softening of the spectrum would normally be explained by
a transition from the usual truncated accretion disk, with optically thin inner flow, to an optically thick flow passing through a boundary layer, as is commonly observed in other LMXBs \citep[see, e.g.,][]{barret02}. However, for \src\ in 2014 June, this transition is not supported by the data, since the optical depths $\tau_1$, $\tau_2$ for both components are {\it lower} than in the hard state. Some caution is required in interpreting these parameters alone, as they are strongly anticorrelated with the corresponding electron temperatures $kT_{e,1}$, $kT_{e,2}$, and we fix the latter at 20~keV.
Since the evidence for increased mass accretion rate is weak, we further attribute the 
markedly different burst behavior,
also to the change in disk geometry.
Although it is presently not understood precisely how the disk geometry can affect the burst behavior, the manifestly different burst properties in the soft state has been observed in several other sources \citep[e.g.][]{cor03}, and this interaction is increasingly being explored in the literature \citep[see, e.g.,][]{Worpel, Kajava, Ji15}, also in the hard state \citep[see, e.g.,][]{intZ12, intZ13}.

\section{Acknowledgements} 
JC would like to thank Niels J\o rgen Westergaard for useful discussions. 
JC acknowledges financial support from ESA/PRODEX Nr. 90057.  
PR acknowledges financial contribution from contract ASI-INAF I/004/11/0 and ASI-INAF I/037/12/0. 
This work made use of data from the \N\ mission, a project led by the 
California Institute of Technology, managed by the Jet Propulsion Laboratory, 
and funded by the National Aeronautics and Space Administration. 
We thank the \N\ and \Sw\ Operations teams for executing the ToO observations, 
and the Software and Calibration teams for analysis support. 
This research has used the \N\ Data Analysis Software
(NuSTARDAS) jointly developed by the ASI Science Data Center (ASDC, Italy) and the California 
Institute of Technology (USA).
The {\it MAXI} data are provided by RIKEN, JAXA and the MAXI team.
\Sw/BAT transient monitor results are provided by the \Sw/BAT team.
This work made use of data supplied by the UK Swift Science Data Centre at the University of Leicester. 
This paper utilizes preliminary analysis results from the Multi-INstrument Burst ARchive (MINBAR), which is supported under the Australian Academy of Science's Scientific Visits to Europe program, and the Australian Research Council's Discovery Projects and Future Fellowship funding schemes.

\smallskip
{\it Facilities:} 
\facility{NuSTAR},
\facility{\Sw},
\facility{MAXI}

\smallskip
\copyright 2014.  All rights reserved.

\end{document}